\documentclass[aps,groupedaddress,showpacs,%
amsmath,preprintnumbers,12pt]{revtex4}
\input{amssym}
\usepackage{hyperref}
\def\p {\partial}
                                                                                
\def\t {\tilde}
                                                                                
\def\be {\begin{equation}}
                                                                                
\def\ee  {\end{equation}}
                                                                                
\def\bea {\begin{eqnarray}}
                                                                                
\def\eea {\end{eqnarray}}
                                                                                
\def\nn {\nonumber}
\begin{document}
\preprint{gr-qc/0503031}
\title{Flat slice Hamiltonian formalism for dynamical black holes}
\author{Viqar Husain and Oliver Winkler}
\email[]{husain@math.unb.ca, oliver@math.unb.ca}
\affiliation{Department of Mathematics and Statistics,
University of New Brunswick, Fredericton, NB, Canada E3B 5A3}
\date{\today}
\begin{abstract}
                                                                                
We give a Hamiltonian analysis of the asymptotically flat spherically 
symmetric system of gravity coupled to a scalar field. This 1+1 dimensional 
field theory may be viewed as the "standard model" for studying black 
hole physics. Our analysis is adapted to the flat slice Painleve-Gullstrand 
coordinates. We give a Hamiltonian action principle for this system, which 
yields an asymptotic mass formula. We then perform a time gauge fixing  
that gives a Hamiltonian as the integral of a local density. The  
Hamiltonian takes a relatively simple form compared to earlier work in 
Schwarzschild gauge, and therefore provides a setting amenable to full 
quantisation. 

\end{abstract}
                                                                                
\maketitle
                                                                                
\section{Introduction} 
It may be argued that the central unsolved problem in black hole
physics is a full quantization of the gravity-scalar field system in
spherical symmetry.  All the puzzles associated with black holes
originated from studying this system classically and semi-classically
\cite{hawk,unruh,bd,rw}. The extra step to full quantization has never
been attempted, mainly due to the intractability of the Hamiltonian
system. If this could be accomplished, we would have a complete
scenario for studying the formation and evaporation of Schwarzschild
black holes in a fully quantum dynamical setting, and in the simplest
"no frills" context. 

Although there has been some progress in this area in string theory,
it is restricted to highly special black holes in derived supergravity
models \cite{stringbh}. In the loop quantum gravity program, the work
on this problem has so far been restricted to "isolated" horizons
\cite{isoh}. As the name suggests, these horizons are not appropriate
for studying the full dynamics of matter and gravity, where horizons
can form and evolve in response to matter flows.

The first attempt at starting a quantization of the gravity-scalar
field model in four spacetime dimensions dates back to the seventies,
when the Hamiltonian theory was worked out in a parametrization
adapted to Schwarzschild coordinates by Berger et. al. (BCNM)\cite{BCMN}, 
and later clarified by Unruh \cite{unruh}. In this setting a reduced
Hamiltonian for the scalar field was obtained in a particular gauge.
This Hamiltonian was complicated enough that quantization was
effectively untenable.  (Unruh comments on this by saying "I present
it here in the hope that someone else may be able to do something with
it." \cite{unruh}). The problem has since remained largely unaddressed, 
except for related work on shell collapse \cite{hajicek,jorma}, and a
geometrodynamical quantization of the (vacuum) Schwarzschild black hole
\cite{kuchar,madhav}.  

In this paper we reanalyse the spherically symmetric gravity-scalar
field system. The main new ingredient is the application of the
Arnowitt-Deser-Misner (ADM) Hamiltonian formulation to coordinates
adapted to the flat slice Painleve-Gullstrand (PG) coordinates 
\cite{kr-wil,jorma,gm,mp,hqs}, with a time gauge fixing. 
The black hole metric in these coordinates is given by
\be 
 ds^2 = -dt^2 + \left(dr + \sqrt{\frac{2M}{r}}\ dt\right)^2 + r^2 d\Omega^2. 
\label{PGm}
\ee
The spatial metric $e_{ab}$ given by the constant $t$ slices is flat,
and the extrinsic curvature of the slices is 
\be K_{ab} =
-\sqrt{\frac{2M}{r^3}}\left( e_{ab} - {\frac{3}{2}}\ s_a s_b\right),
\label{K}
\ee
where $s^a=x^a/r$ for Cartesian coordinates $x^a$ \cite{flat}.  

The black hole mass information is contained only in the extrinsic curvature, which 
in the canonical ADM variables $(q_{ab},\tilde{\pi}^{ab})$ determines the
momenta $\tilde{\pi}^{ab}$ conjugate to the spatial metric
$q_{ab}$. In this form, the mass formula is necessarily different from 
the ADM mass integral, since the spatial slices are flat. 

The PG coordinates motivate the following prescription for the 
falloff conditions  for asymptotic flatness: 
\bea
q_{ab} &=& e_{ab} + \frac{f_{ab}(\theta,\phi)}{r^{3/2+\epsilon}} 
+ {\cal O}\left(r^{-2}\right)  \label{pgq}\\
\pi^{ab} &=& \frac{g^{ab}(\theta,\phi)}{r^{3/2}} 
+  {\cal O}\left(r^{-3/2-\epsilon}\right),
\label{pgpi}
\eea
where $\epsilon>0$, $f^{ab},g^{ab},h^{ab}$ are symmetric tensors,
$\pi^{ab}=\t\pi^{ab}/\sqrt{q}$, and $q={\rm det}(q_{ab})$. In this
definition it is manifest that the leading terms correspond to the
black hole solution in PG coordinates. The form of the next terms 
containing $\epsilon$ is necessitated by the requirement of a well
defined action principle. It is useful to compare these with the
conditions motivated by the Schwarzschild coordinates, where the leading
order terms in the metric and extrinsic curvature are $1/r$ and
$1/r^2$, respectively \cite{rt,bm}.

These falloff conditions provide the starting point for our analysis of
the Hamiltonian dynamics for spherically symmetric metrics of the form 
\be 
ds^2 = -f(r,t)^2dt^2 + \left(dr + g(r,t)dt\right)^2 + r^2 d\Omega^2, 
\ee
minimally coupled to a massless scalar field $\phi(r,t)$. 

In the next section, we give a parametrization for the ADM variables
$(q_{ab},\pi^{ab})$ respecting these conditions to obtain a
Hamiltonian theory. In Section III, we utilise a special time gauge
fixing condition in which the reduced Hamiltonian takes a relatively
simple form compared to the earlier works mentioned above. The setting
therefore provides an arena in which a full quantization appears to be
possible. The last section contains a brief comparison with the BCNM
work, and a discussion of our approach to quantization. The present
paper also provides the classical details which underlie the recent work on
quantization by the authors \cite{hw1,hw2}.
                                                                                                 
\section{Gravity-scalar field model}

A well defined variational principle for Einstein's equations coupled
to matter, and satisfying a specified class of boundary conditions,
may be obtained by starting with the bulk Einstein-Hilbert action, or
its canonical bulk ADM form.  The first requirement is that all the
terms in the action are well-defined.  The second requirement is that
the variation of the action be of a form such that all variational
derivatives with respect to the field variables are well defined. 
As noted by Regge and Teitelboim \cite{rt,bm}, this requires in general 
the addition of a surface term to the original action.

The phase space of the model is defined by prescribing a
form of the gravitational phase space variables $q_{ab}$ and
$\tilde{\pi}^{ab}$, together with falloff conditions in $r$ for these
variables, and for the lapse and shift functions $N$ and $N^a$, such
that the ADM 3+1 action for general relativity minimally coupled to a
massless scalar field
\be
S = \frac{1}{16\pi G}\int d^3x dt\left[ \tilde{\pi}^{ab}\dot{q}_{ab} +
\t{P}_\phi\dot{\phi}
- N H - N^a C_a\right]
\label{act}
\ee
is well defined. The constraints arising from varying the lapse and shift
are
\bea
{\cal H} &=& \frac{1}{\sqrt{q}}\left(\tilde{\pi}^{ab}\tilde{\pi}_{ab}
-\frac{1}{2} \tilde{\pi}^2 \right)
                       \sqrt{q}R(q) \nn\\
          & &- 8\pi G \left(\frac{1}{\sqrt{q}}\tilde{P}_\phi^2
                       + \sqrt{q}q^{ab}\p_a\phi\p_b\phi\right) = 0  \\
{\cal C}_a &=& D_c\t{\pi}^c_a - 8\pi G\t{P}_\phi\p_a\phi =0,
\eea
where $\t{\pi}=\t{\pi}^{ab}q_{ab}$. The corresponding conditions for
the matter fields $\phi$ and $\t{P}_\phi$ are determined by the
constraint equations.
 
In this setting, the following parametrization for the 3-metric and
conjugate momentum gives a reduction to spherical symmetry: 
\bea
q_{ab} &=& \Lambda(r,t)^2\ s_a s_b + \frac{R(r,t)^2}{r^2}\ ( e_{ab} - s_a
s_b)\\
\t{\pi}^{ab} &=& \frac{P_\Lambda(r,t)}{2\Lambda(r,t)}\ s^a s^b + \frac{r^2
P_R(r,t)}{4R(r,t)}\
(e^{ab} - s^a s^b).
\label{reduc}
\eea

Substituting these into the 3+1 ADM action shows that the pairs $(R,P_R)$
and $(\Lambda,P_\Lambda)$ are canonically conjugate variables. The reduced ADM 1+1 
field theory action takes the form
\bea
S_R &=& \frac{1}{4G}\int dtdr \left(P_R\dot{R} + P_\Lambda\dot{\Lambda} +
P_\phi\dot{\phi}
    \right)\nn\\
     && -\frac{1}{4G}\int dtdr \left( NH + N^r C_r\right) \nn\\
 && +\ {\rm surface\ term},
\eea
where we have performed the angular integral. The surface term is derived below. 
The Hamiltonian and diffeomorphism constraints are 
\bea
H &=& \frac{1}{R^2\Lambda}\left[\frac{1}{8}\ (P_\Lambda \Lambda)^2 -
\frac{1}{4}(P_\Lambda \Lambda)(P_R R)\right]
\nn\\
&&+ \frac{2}{\Lambda^2}\left[ 2RR''\Lambda -2RR'\Lambda' - \Lambda^3 +
\Lambda R'^2 \right] \label{Hsph}\nn\\
&& + \left[\frac{P_\phi^2}{2\Lambda R^2} + \frac{R^2}{2\Lambda}\ \phi'^2
\right].\\
C_r &=&   P_R R'  -\Lambda P_\Lambda' + P_\phi\phi' = 0.
\label{Csph}
\eea
These constraints are first class. The falloff conditions induced on the
reduced variables by (\ref{pgq}-\ref{pgpi}) are 
\bea
R &=& r +{\cal O}(r^{-1/2-\epsilon}),\nn \\
P_R &=& Ar^{-1/2}/2 + {\cal O}(r^{-1-\epsilon}), \label{R}\\
\Lambda &=& 1 + {\cal O}(r^{-3/2-\epsilon}),\nn \\ 
P_\Lambda &=& A r^{1/2} + {\cal O}(r^{-\epsilon}) \label{Lam}\\
\phi &=& Br^{-1/2} + {\cal O}(r^{-3/2-\epsilon}),\nn \\ 
P_\phi &=& Cr^{1/2} + {\cal O}(r^{-\epsilon}),
\label{phi}
\eea
where $A,B,C$ are constants. This means that the asymptotic region is 
not dynamical (as it should be since it is flat). The constant $A$, which appears 
in the expressions for $P_R$ and $P_\Lambda$ above, will turn out to be captured 
by a surface integral (see below) and is proportional to the mass of the system. 
The above conditions together with the falloff conditions on the lapse and shift 
functions
\bea
N^r &=& Ar^{-1/2} + {\cal O}(r^{-1/2-\epsilon}) \nn\\
N &=& 1 + {\cal O}(r^{-\epsilon})
\label{asympN}
\eea
ensure that $S_R$ is well-defined. More explicitly, they ensure that
the symplectic structure is well-defined, which means that the
integral of the terms $P_R\dot{R}$ etc. converges. Furthermore, they
guarantee that $H$ vanishes to first order, and has a falloff
$r^{-1-\epsilon}$ beyond leading order, as required for the action to
be well-defined. The same holds for the diffeomorphism constraint.
Note that the factor $1/2$ in the leading order term in $P_R$
ensures that to this order both the diffeomorphism and Hamiltonian
constraints vanish. This corresponds to the black hole solution in PG
coordinates, which motivated our falloff conditions in the first
place. Taken together these observations guarantee that the bulk
action is well-defined.
               
Consider now the variation of this action to see what surface terms
need to be added to make the variational principle
well-defined. Following \cite{rt}, we compute the variation $\delta
S_R$, see what surface terms arise in it, and identify the terms that
vanish due to the falloff conditions; the ones that do not must
be compensated for by adding a surface term to the starting bulk
action. In our case surface terms arise from those terms in the action that
contain $r$ derivatives. These are the Ricci scalar, matter density,
and the radial diffeomorphism terms. The variation is
\bea
4G\ \delta S_R &=& \int dt dr\ \left( {\rm terms\ giving\ eqns.\ of\ motion} \right)\nn\\ 
&& -\int  dt \left[N^r P_\phi + \frac{NR^2\phi'}{\Lambda}\right]\delta \phi   \nn\\ 
&&+\int dt \left[N^r \Lambda\ \delta P_\Lambda + \frac{4 N RR'}{\Lambda^2}\
\delta\Lambda\right]  \nn \\ 
&& +\int dt\left[8\left(\frac{NR}{\Lambda}\right)' - \frac{4NR'}{\Lambda}
+\frac{4NR\Lambda'}{\Lambda^2} \right] \delta R  \nn\\ 
&& -\int  dt\left[ N^r P_R\right] \delta R,
\eea
where the last four terms are differences of surface integrals evaluated 
at $r=0$ and $r=\infty$. The variational principle is well-defined if each of 
these surface terms vanishes. 

At $r=\infty$ most of these vanish by virtue of the falloff 
conditions. This leaves two terms of order one. The one proportional to $\delta\phi$
can be eliminated by requiring this variation to vanish at infinity. The other, 
proportional to $\delta P_\Lambda$, is dealt with by  adding a surface term at infinity 
to the original bulk action whose variation cancels the offending surface term from the
variation. It is this term that captures the conserved asymptotic mass.

At $r=0$ we do not impose falloff conditions on the phase space variables because 
there is no physical guidance for this. We thus require  the usual prescription that 
the variations of the configuration variables vanish there:  
\bea
\delta\phi  |_{r=0} &=& 0,\nn\\
\delta R |_{r=0} &=& 0,\nn\\
\delta \Lambda |_{r=0} &=& 0.
\eea 
This leaves only the term proportional to $\delta P_\Lambda$ at $r=0$. 
To deal with this we require either the addition of a surface term 
with the opposite sign to the one at infinity, or the condition $N^r(r=0)=0$. 
The former would subtract from the asymptotic mass. Therefore for the vacuum solutions 
($\phi=0$ and $P_\phi=0$) in the gauge $R=r$ $\Lambda=1$, the sum of the surface terms 
at $r=0$ and $r=\infty$ would cancel. This suggests that we make the latter choice. 

Based on these observations, functional
differentiability of the action is guaranteed if we add the term
\be
 - \int dt \left(N^r \Lambda P_\Lambda\right)|_{r=\infty}
\label{surterm}
\ee
to the original bulk action. We can then derive the evolution equations: 
\bea
\dot{R} &=& -N \frac{P_\Lambda}{4R} + N^rR'\label{evolR}\\
\dot{P_R} &=& N\left[ \frac{P_\Lambda^2 \Lambda}{4R^3} - \frac{P_RP_\Lambda}{4R^2}
+ \frac{P_\phi^2}{\Lambda R^3}
-\frac{R}{\Lambda}\phi'^2\right]\nn \\
&&- \left(\frac{4R\Lambda' N}{\Lambda^2}\right)' + \frac{4R'N'}{\Lambda} 
- \left(\frac{4RN}{\Lambda}\right)' \nn \\ 
&& + (N^r P_R)'\\
\dot{\Lambda} &=& \frac{N}{4R^2\Lambda}\left( P_\Lambda \Lambda^2 -\Lambda RP_R\right) 
+ (\Lambda N^r)'\label{Ldot}\\
\dot{P}_\Lambda &=& N\left( -\frac{P_\Lambda^2}{8R^2} + \frac{4RR''}{\Lambda^2} +2 
+ \frac{2(R')^2}{\Lambda^2}
-\frac{8RR'\Lambda'}{\Lambda^3}\right)\nn\\
&& - \left( \frac{4RR'N}{\Lambda^2}\right)' 
+ \frac{N}{2\Lambda^2R^2}\left(P_\phi^2 +R^4(\phi')^2\right) \nn \\
&& + N^rP_\Lambda'\\
\dot{\phi} &=& N \frac{P_\phi}{\Lambda R^2} + N^r \phi' \\
\dot{P_\phi} &=&  \left( N\ \frac{\phi' R^2}{\Lambda} \right)' + (N^r P_\phi)'.
\label{evolphi}
\eea

Note that the surface term (\ref{surterm}) is the mass formula for the flat slice
parametrization. Substituting the asymptotic forms of the variables (\ref{Lam})
and (\ref{asympN}) gives  
\be 
   N^r\Lambda P_\Lambda = A^2 + {\cal O}(r^{-\epsilon}), 
\ee
which shows that the parameter $A$ contains conserved mass information. The relation
between $A$ and the conventional mass parameter $M$ in (\ref{PGm}) is obtained by 
comparing the $\pi^{ab}$ obtained from (\ref{K}) with our parametrization (\ref{reduc}). 
This gives $A=4\sqrt{2M}$. We note also that our falloff conditions are preserved under 
this evolution, which ensures their consistency. This is easily seen by computing the 
left and right hand sides of the above evolution equations in the asymptotic regime.   

\section{Time Gauge Fixing}
                                                                                
We now gauge fix the theory defined above with the condition
$\Lambda = 1$. It is second class with the Hamiltonian
constraint. Demanding that it be preserved in time implies from 
(\ref{Ldot}) the relation
\be
N(P_\Lambda - RP_R) = -4R^2 (N^r)'
\ee
between the lapse $N$ and the shift $N^r$. As a result of the time gauge
fixing the Hamiltonian constraint (\ref{Hsph}) must be imposed strongly. This gives 
a quadratic equation  for $P_\Lambda$ in terms of the remaining
variables. A comparison with the vacuum solution in fully gauge-fixed form 
(ie. with the coordinate fixing condition $R=r$), uniqely selects the positive 
root. This gives
\be
P_\Lambda = P_RR + \sqrt{ (P_RR)^2 - X},
\label{PL}
\ee
where
\be
X =  16R^2 (2RR'' - 1 + R'^2) + 16R^2 H_\phi
\ee
and
\be
 H_\phi = \frac{P_\phi^2}{2R^2} + \frac{R^2}{2}\ \phi'^2.
\ee
The positivity of the argument of the square root follows from the dominant energy 
condition for the massless scalar field.    

The solution for the lapse function now reads
\be
N = -\frac{4R^2 (N^r)'}{\sqrt{(P_RR)^2 - X}}.
\label{lapse}
\ee
                                                                                
The reduced Hamiltonian equations for the remaining canonical
variables $(R,P_R)$ and $(\phi,P_\phi)$ are obtained by substituting
the gauge condition $\Lambda=1$, the corresponding solution (\ref{PL}) of the
Hamiltonian constraint, and the lapse equation
(\ref{lapse}) into the unfixed evolution equations (\ref{evolR})-(\ref{evolphi}), 
and into the radial diffeomorphism constraint.  
The gauge fixed equations are  
\bea 
\dot{R} &=& -\frac{N}{4R}\ \left(P_RR + \sqrt{(P_RR)^2 - X}\right) + N^rR'\\
\dot{P_R} &=& \frac{N}{4R^3}\ \left( P_R^2R^2 + P_R R\sqrt{(P_RR)^2 - X} - X \right)\nn\\ 
&& + N\left(\frac{P_\phi^2}{R^3} - R \phi'^2\right)+ 4R'N' - \left(4RN\right)' \nn\\ 
&& + (N^r P_R)'\\
\dot{\phi} &=& N\ \frac{P_\phi}{R^2} + N^r \phi' \\
\dot{P_\phi} &=&  \left( N \phi' R^2 \right)' + (N^r P_\phi)',
\eea
where it is understood that $N$ is given by (\ref{lapse}). The remaining 
radial diffeomorphism constraint is 
\be 
P_\Lambda' + P_RR' + P_\phi \phi' = 0,
\label{remdiff}
\ee 
where $P_\Lambda$ is given by (\ref{PL}).

All these equations can be obtained from the gauge fixed reduced action
\bea 
S_R^G &=& \int dt dr \left[P_\phi\dot{\phi} + P_R\dot{R} - N^r(P_\Lambda' + P_RR' + P_\phi \phi')\right]\nn\\
&& + \int dt P_\Lambda (N^r)'.
\eea
The surface term may be combined with the bulk term to write the action in a form 
from which one can read off the gauge fixed Hamiltonian 
\bea 
H_{R}^G &=& \int_0^\infty \left[(N^r)'P_\Lambda + N^r(P_RR' + P_\phi \phi')\right]dr \nn\\
&=&\int_0^\infty (N^r)'\left( R P_R + \sqrt{(P_RR)^2 - X}\right) dr\nn\\
  && + \int_0^\infty N^r(P_RR' + P_\phi \phi')\ dr,
\label{Hred}
\eea

\section{Discussion} 

The Hamiltonian (\ref{Hred}) is a simpler expression than that obtained 
from the full time and coordinate gauge fixing given in \cite{unruh}, where 
Schwarzschild gauge is used. That Hamiltonian is 
\be 
 H_S = \int dr\left(\frac{P_\phi^2}{4r^2} +r^2 \phi'^2\right)\exp\left(\int_\infty^r S_\phi(r')dr'\right) 
\label{Hu}
\ee
where 
\be 
S_\phi(r) = \frac{P_\phi^2}{8r^3} + \frac{r\phi'^2}{2}.
\ee
It is apparent that the square root in (\ref{Hred}) is easier to handle than the 
non-locality manifest in this formula. 

The non-local term can be traced back to the fact that (\ref{Hu}) is arrived at after 
both a time and radial coordinate gauge fixing. This may be seen in our 
formulation as well: Consider the radial gauge fixing $R(r,t)=r$. This  
leads to (i) fixing of the shift function $N^r$ by the condition  
\be 
 \dot{r} = 0 = (N^r)' r \ \left(1+ \frac{rP_R }{ \sqrt{(rP_R)^2 - X}}\right) + N^r\\
\label{Nr}
\ee  
and (ii) fixing of $P_R$ by strong imposition of the remaining diffeomorphism constraint 
(\ref{remdiff}).  The solution of (\ref{Nr}) for $N^r$ is 
\be 
 N^r = \exp\left(- \int_\infty^r \frac{\sqrt{(\bar{r}P_R)^2 - X}}{\bar{r}\sqrt{(\bar{r}P_R)^2 - X}
+ \bar{r}^2P_R}\ d\bar{r}\right),
\ee
which  leads directly to a non-local term when substituted back into (\ref{Hred}). Thus we 
learn that the complicated form of the fully gauge fixed Hamiltonian is traceable solely to 
the coordinate gauge fixing, but not the time gauge fixing. A partial gauge fixing 
in the diagonal parametrization of the metric \cite{BCMN,unruh} would also lead to a 
simpler, local form of the reduced Hamiltonian.

For this reason, we propose that a quantization of this system be carried out in a
partially gauge fixed setting, with only the time gauge fixed. Retaining a first class 
radial diffeomorphism constraint presents issues that are relatively easier to deal with; 
we have in fact already proposed \cite{hw1} a kinematical quantization of this system 
in which the Hilbert space carries a representation of finite diffeomorphisms.  
 
In summary, we have presented here in detail the canonical formalism for spherically symmetric 
gravity coupled to a scalar field adapted to the flat slice foliation. We have shown that 
all consistency conditions are satisfied. These include functional differentiability, 
surface terms, and the preservation under evolution of the falloff conditions at infinity. 

Having obtained the reduced Hamiltonian (\ref{Hred}), the next step in our program is to construct 
the corresponding operator using the techniques presented in \cite{hw1}. This work 
is to appear \cite{hw3}. 

\acknowledgments{We are grateful to Jorma Louko for illuminating discussions and comments on 
an earlier draft of the paper. This work was supported in part by the Natural Science and Engineering Research Council of Canada.}


\begin{thebibliography}{99}
\bibitem{flat} The flat slice foliation covers the most interesting regions of spacetimes describing 
realistic matter collapse. As such it is especially suited for studying Hawking radiation 
from a fully quantum gravitational viewpoint, which is the central motivation for our work. 
Indeed, since no "extended" spacetimes are known for dynamical matter collapse, the 
long time static Schwarzschild limit with its full Kruskal extension is an 
"isolated" point in the solution space for generic spherically symmetric collapse.
\end{thebibliography}
\end{document}